\documentstyle[prl,aps,epsf]{revtex}
\begin{document}
\draft
\title{Energy dissipation statistics in a shell model of turbulence}

\author{G. Boffetta$^{1}$, A. Celani$^{1,2}$ and  D. Roagna$^{1}$}
\address{$^{1}$ Dipartimento di Fisica Generale, Universit\`a di Torino,
	Via Pietro Giuria 1, 10125 Torino \\
	and INFM Unit\`a di Torino Universit\`a, Italy}
\address{$^{2}$ Observatoire de la C\^ote d'Azur, BP 4229, F-06304 Nice Cedex 04, France}
\date{\today}

\maketitle

\begin{abstract}
The Reynolds number dependence of the statistics of energy
dissipation is investigated in a shell model of fully developed
turbulence. 
The results are in agreement with a model which 
accounts for fluctuations of the dissipative scale with
the intensity of energy dissipation. 
It is shown that the assumption of a fixed dissipative scale 
leads to a different scaling with Reynolds which is not compatible 
with numerical results. 
\end{abstract}

\pacs{}

One of the most important problems in fully developed turbulence
is the description of the energy transfer mechanism. 
In stationary situations, the energy injected at
the large scale $\ell_0$, transfers at rate $\bar{\epsilon}$ down to
the dissipative scale $\eta$, where it is removed at the same rate
by viscous dissipation. The fundamental assumption in the study
of fully developed turbulence is that in the limit of very
high Reynolds numbers $Re$, the energy dissipation $\bar{\epsilon}$
becomes independent of $Re$ (i.e. of the viscosity, being
$Re = u_0 \ell_0 /\nu$, with $u_0$ a typical large scale velocity)
\cite{MY75,Frisch95,Sreenivasan84,Sreenivasan98}.
In the same limit, the Kolmogorov theory predicts
universal scaling of the velocity structure functions 
in the inertial range of scales $\eta < \ell < \ell_0$:
\begin{equation}
S_q(\ell) \equiv \langle \left(\delta u(\ell)\right)^q
\rangle \sim u_0^q \left( {\ell \over \ell_0} \right)^{\zeta_q}
\label{eq:1}
\end{equation}
with exponents $\zeta_q=q/3$.

Several decades of experimental and numerical investigation have
shown that scaling laws (\ref{eq:1}) are indeed observed
but with exponents $\zeta_q$ corrected with respect to the 
Kolmogorov prediction \cite{AGHA84}. This is the essence of the
intermittency problem, which has received a lot of attention
in the modern approach of the study of fully developed turbulence.

Experiments have shown that intermittency also affects energy dissipation
statistics \cite{MS87} which is not uniform in the turbulent domain.
A phenomenological description of intermittency is
the multifractal model \cite{PF85}.
This model introduces a continuous set of scaling exponents $h$ which relate
the velocity fluctuations entering in (\ref{eq:1}) with the large
scale velocity $u_0$: 
\begin{equation}
\delta u(\ell) \sim u_0 \left({\ell \over \ell_0}\right)^h  \, .
\label{eq:1.2}
\end{equation}
The exponent $h$ is realized with a probability 
$\left({\ell \over \ell_0}\right)^{Z(h)}$ where $Z(h)$ is the 
codimension of the fractal set on which the $h$-scaling holds.
The scaling exponents of structure functions (\ref{eq:1}) are 
obtained by a steepest descent argument over exponents $h$:
\begin{equation}
\zeta_q = \inf_{h} \left[p h + Z(h) \right] \, .
\label{eq:1.3}
\end{equation}

The scaling region is bounded from below by the 
Kolmogorov dissipation scale $\eta$ at which dissipation starts
to dominate, i.e. the local Reynolds number is of order 1:
\begin{equation}
{\eta \delta u(\eta) \over \nu} \sim 1
\label{eq:2}
\end{equation}
At variance with the Kolmogorov theory, in the multifractal 
description of intermittent turbulence, the dissipative scale 
is a fluctuating quantity.
This implies a series of consequences which have been investigated
in past years \cite{PV87,FV91}. As we shall see later 
the description of the fluctuations of the dissipative scale 
is crucial for the correct evaluation of the Reynolds number 
dependence.

In this Paper we are interested in the dependence of
the statistics of energy dissipation on the Reynolds number. 
The physical picture is that dissipation becomes more and
more intermittent as the Reynolds number increases. 
Assuming that the multifractal description can be pushed down 
to the dissipative scale, one predicts for the moment of energy 
dissipation a power-law dependence on $Re$, with exponents related 
to the structure function exponents (\ref{eq:1}) \cite{Nelkin90,Frisch95}.
We will see that this prediction is rather natural and confirmed 
by numerical simulations on a shell model. 

The dimensional argument for the prediction goes as follows. In
a dimensional approach, the energy dissipation is evaluated as
\begin{equation}
\epsilon = \nu \sum_{\alpha,\beta}\left(\partial u_{\alpha} \over 
\partial x_{\beta} \right)^2 \sim
\nu \left( \delta u(\eta) \over \eta \right)^2
\label{eq:3}
\end{equation}
 From (\ref{eq:1.2}) and (\ref{eq:2}) one has that 
$\eta \sim \ell_0 Re^{-{1 \over 1+h}}$. 
Inserting in (\ref{eq:3}) and computing
the average of the different moments, one ends with the expression
\begin{equation}
\langle \epsilon^p \rangle \sim \bar{\epsilon} \, ^p \int d \mu(h) \, 
Re^{-{3 p h - p +Z(h) \over 1+h}} \sim \bar{\epsilon} \, ^p Re^{-\theta_p}
\label{eq:4}
\end{equation}
where the integral has been evaluated by a steepest descent 
argument (assuming $Re \to \infty$) and
\begin{equation}
\theta_p = \inf_{h} \left[{3 p h - p + Z(h) \over 1 + h}\right] \, .
\label{eq:5}
\end{equation}

The standard inequality in the multifractal model (following 
from the exact result $\zeta_3=1$), $Z(h) \ge 1-3h$,
implies for (\ref{eq:5}) $\theta(1)=0$ which is nothing but the
request of finite nonvanishing dissipation in the limit $Re \to \infty$.
For $p > 1$, $\theta_p < 0$, i.e. the tail of the distribution
of $\epsilon$ becomes wider with Reynolds number.

Let us stress that the above argument is only a reasonable dimensional 
argument. It is essentially based on two assumptions: a physical one
concerning the fluctuations of the dissipative scale according to (\ref{eq:2}),
and a more formal one on the possibility of extending the
multifractal description down to the dissipative scales.
The two assumption are independent: indeed, as we will see, it is possible to
give different predictions by changing assumption (\ref{eq:2})
\cite{EG92}.

It would thus be important to address the problem with experiments 
or direct numerical simulation at high Reynolds numbers. Recent
high resolution DNS gives support to (\ref{eq:5}) \cite{CB99}, 
but the Reynolds number is not large enough to discriminate 
clearly between different predictions. 

Shell models are extremely simplified models of turbulence.
Nevertheless, they are deterministic, nonlinear dynamical models 
which display intermittency and anomalous scaling exponents
reminiscent of real turbulence \cite{BJPV98}.
Their main advantage is that with shell models one can perform 
very accurate simulations at very high Reynolds numbers; for this
reason they are thus natural candidates for a numerical investigation
of Reynolds number dependence.

In shell models, the velocity fluctuations are represented 
by a single complex variable $u_n$ on shells
of geometrically spaced wavenumber $k_n=k_0 \, 2^n$.
The particular model we adopt for
our investigation is a recently introduced model which displays 
strong intermittency corrections \cite{sabra}.  
The model equations are
\begin{equation}
\frac{d u_n}{dt}=i k_n \left( u_{n+2}u_{n+1}^*
 -\frac{1}{4} u_{n+1}u_{n-1}^*+\frac{1}{8} u_{n-1}u_{n-2} \right)  
-\nu k_n^2  u_n +f_n
\label{eq:6}
\end{equation}
where $\nu$ is the viscosity and $f_n$ is a forcing term restricted
to the first two shells. For $\nu=f_n=0$ the model conserves the 
total energy $E=\sum_n |u_n|^2$. 
For simplicity, the forcing adopted for the present simulations
is $f_n \propto 1/u_n^{*}$, which guarantees a constant energy input
$\bar{\epsilon}$.
The large scale Reynolds number of the simulation is 
estimated as $Re = \bar{\epsilon}^{1/3}/(\nu k_0^{4/3})$ and is 
numerically controlled by the value of the viscosity.

The chaotic dynamics is responsible for intermittency corrections to 
the structure functions exponent $\zeta_q$, here defined by means of
\begin{equation}
S_q(n) = \langle |u_n|^q \rangle \sim k_n^{-\zeta_q} \, ,
\label{eq:7}
\end{equation}
which are close to the experimental values \cite{BJPV98}.
In Figure~\ref{fig1} we plot the spectrum of structure function
exponents obtained from very long simulations. The multifractal
codimension $Z(h)$ is numerically obtained from $\zeta(p)$ by 
inverting the Legendre transform (\ref{eq:1.3}).
The result is shown in Figure~\ref{fig2}. 
We observe that, because of the strong intermittency
in the model, it is numerically difficult to obtain statistical
convergence of structure functions (\ref{eq:7}) of order $q>8$.
As a consequence, the minimum exponent for $Z(h)$ is $h_{min} \simeq 0.2$.

From the energy balance equation we have the instantaneous energy dissipation
\begin{equation}
\epsilon = 2 \nu \sum_{n} k_n^2 |u_n|^2 
\label{eq:8}
\end{equation}
which average is
$\langle \epsilon \rangle = \bar{\epsilon}$ in stationary conditions.

We have performed very long simulations at different Reynolds 
numbers, starting from $Re = 2 \times 10^{5}$ up to $Re = 10^{8}$. 
For each simulation we computed the different moments of
energy dissipation, $\langle \epsilon^p \rangle$.
Shell models dynamics is characterized by strong bursts
of energy dissipation which limits the possibility of
computing with confidence high order moments.
Here we limited to moments  $p \le 8$.
In Figure~\ref{fig3} we plot the behavior of 
$\langle \epsilon^p \rangle$ as a function of $Re$ for 
different values of $p$. The power law behavior is evident
and the scaling exponent $\theta_p$ can be estimated with
good accuracy. By construction 
$\langle \epsilon \rangle=\bar{\epsilon}$ 
is independent of $Re$.

The scaling exponent $\theta_p$ (\ref{eq:4}) are plotted in 
Figure~\ref{fig4}, together with the multifractal prediction
(\ref{eq:5}).
Let us observe that, because the largest $q$ in
(\ref{eq:7}) is $q=8$, the estimate of $p$ in (\ref{eq:5}) 
is limited to values less than $p \simeq 2.5$. 
For higher $p$, the numerical evaluation of 
(\ref{eq:5}) feels the effect of the cutoff of $h$ on $h_{min}$.
Nevertheless, we have a rather large range of moments ($0 \le p \le 2.5$)
over which numerical data display a perfect agreement with 
(\ref{eq:5}).

As discussed above, prediction (\ref{eq:6}) makes use of the
fluctuating dissipative scale $\eta$. If, on the contrary, one
assumes that dissipation scale enters in (\ref{eq:3}) as an
averaged quantity the prediction for $\theta_p$ is different:
assuming 
$\nu\langle\delta u(\tilde{\eta})^2 \rangle/\tilde{\eta}^2 \sim \bar{\epsilon}$ as the definition of the (non-fluctuating) dissipation scale $\tilde{\eta}$ (this
is the only choice that ensures that $\langle \epsilon \rangle\sim Re^0$) , 
one ends up with
$\tilde{\theta}_p=[p\zeta_2-\zeta_{2p}]/[2-\zeta_2]$ \cite{EG92}.

Our results allow us to discriminate between 
the prediction (\ref{eq:5}) 
and the one obtained with a non-fluctuating dissipative scale.
Figure~\ref{fig4} shows that the numerical $\theta_p$  is definitely 
not compatible with the latter alternative,
whereas it supports with good accuracy the prediction (\ref{eq:5}).

In conclusion, it is an expected consequence of the 
existence of intermittency in the energy transfer that
the dissipative scale $\eta$ fluctuates according to the local 
intensity of energy dissipation, being smaller where the dissipation
is stronger and vice versa. The fluctuations of the inner
scale of turbulence reflect onto the Reynolds dependence of
the statistics of energy dissipation. Long  numerical simulations
of shell-models confirm with great accuracy the validity  
of the multifractal model, which accounts for the fluctuations of
$\eta$, and rule out alternative models which do not describe properly 
the correlations between $\eta$ and $\epsilon$.


\newpage

\begin{figure}[ht]
\caption{
Structure function exponents $\zeta_q$ obtained from 
simulation of the Shell Model. The number of
shells is $N=24$ and $\nu=10^{-7}$ corresponding
to $Re = 10^{8}$.
}
\label{fig1}
\end{figure} 

\begin{figure}[ht]
\caption{
$Z(h)$ computed by inverting the Legendre transformation
from the data of Figure~\ref{fig1}.
}
\label{fig2}
\end{figure} 

\begin{figure}[ht]
\caption{ 
Moments of energy dissipation $\langle \epsilon^p \rangle$ as
function of $Re$ for $p=1$ ($+$), $p=2$ ($\times$) and
$p=3$ ($\ast$).
}
\label{fig3}
\end{figure} 

\begin{figure}[ht]
\caption{ 
Energy dissipation scaling exponents $\theta_p$. 
Symbols represent the exponents obtained from the
fit of Figure~\ref{fig3}.
Continuous line is prediction (\ref{eq:5}) taking
into account the fluctuations of the dissipative
scale. Dashed line is the prediction obtained by
assuming an average dissipative scale.
}
\label{fig4}
\end{figure} 

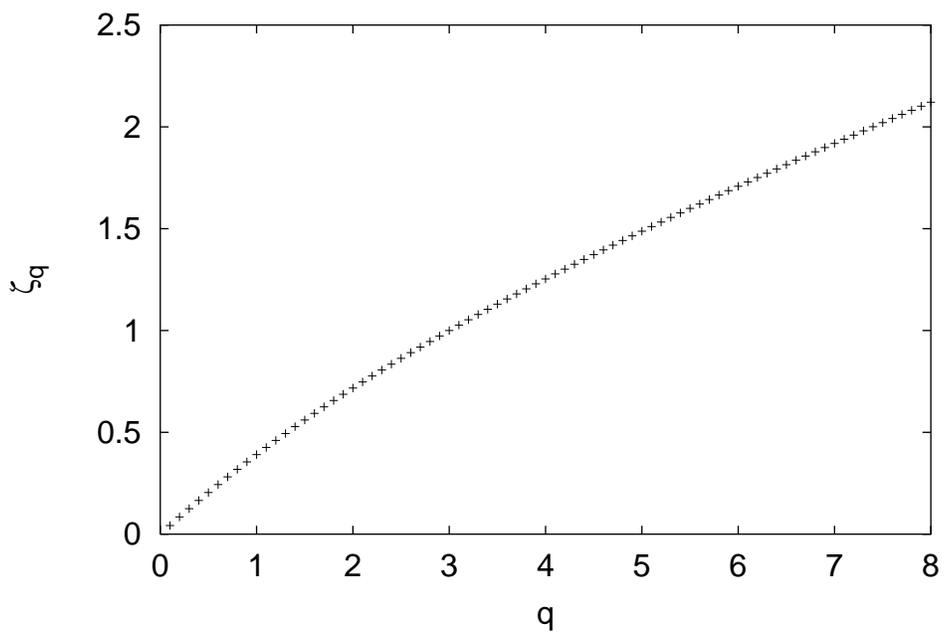
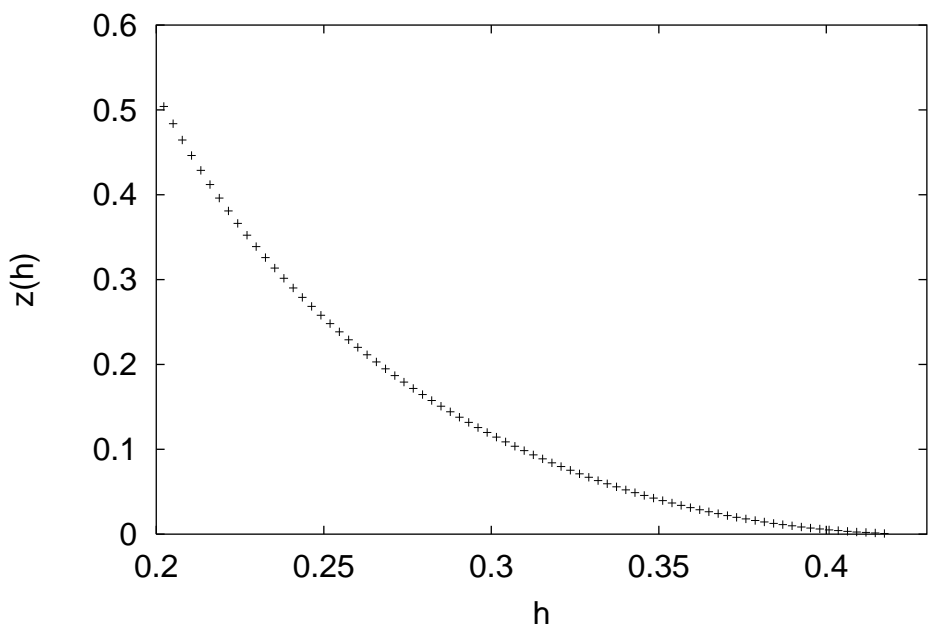
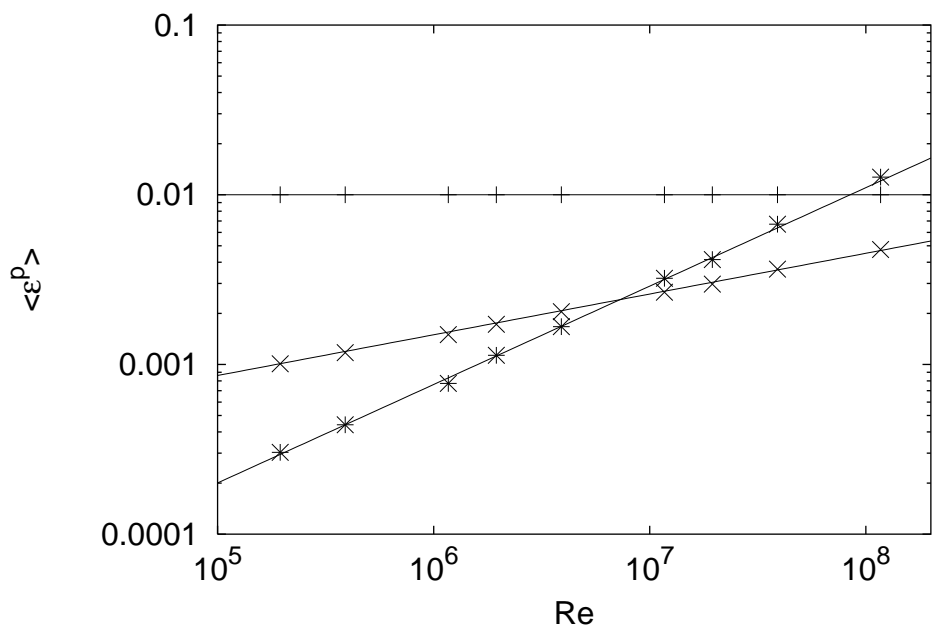
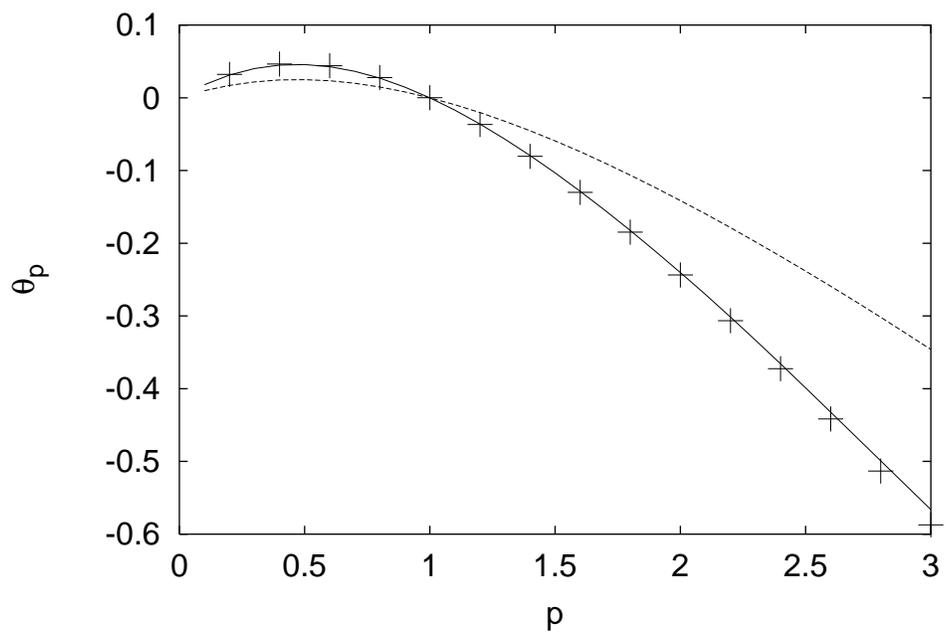
\end{document}